\documentstyle[prl,aps,amsfonts,amssymb,twocolumn,epsfig]{revtex}

\title{
A mechanism for the non-Fermi-liquid behavior in CeCu$_{6-x}$Au$_x$ }
 \author{A. Rosch $^{ 1}$, A. Schr\"{o}der $^{ 2,3}$, O. Stockert $^{ 2}$,
 and H. v. L\"{o}hneysen $^{ 2}$}
\address{$^1$ Institut f\"ur Theorie der Kondensierten Materie, 
Universit\"at Karlsruhe,
D--76128 Karlsruhe, Germany\\
$^2$ Physikalisches Institut, 
Universit\"at Karlsruhe,
D--76128 Karlsruhe, Germany\\
$^3$ Ris\o{} National Laboratory, DK-4000 Roskilde, Denmark}

\begin{document}

\newbox{\dummy}
\savebox{\dummy}{\cite{loehneysenUeb}}
\newcommand{\citeL}{\usebox{\dummy}}

\draft

\twocolumn[\hsize\textwidth\columnwidth\hsize\csname
@twocolumnfalse\endcsname %
\date{\today} \maketitle

\begin{abstract}
  We propose an explanation for the recently observed 
  non-Fermi-liquid behavior of metallic alloys
  CeCu$_{6-x}$Au$_x$: near 
  $x=0.1$, the 
  specific heat $c$ is 
  proportional to $ T \ln (T_0/T)$ and the resistivity increases linearly with
  temperature $T$ over a wide range of $T$.
  These features follow from a model in which
  three-dimensional conduction electrons are coupled to
  {\em two}-dimensional critical {\em ferromagnetic} fluctuations near the
  quantum critical point, $x_{c}=0.1$.
  This picture is motivated by the neutron scattering data in the
  ordered phase ($x=0.2$) and is consistent with the observed
  phase diagram.
\end{abstract}



%
\vskip1.0pc]

Fermi-liquid (FL) theory has traditionally led to an accurate description
of the low temperature properties of metals. Even in
the heavy-fermion compounds, where the bare electron
parameters are renormalized by
 up to three orders of magnitude by the interaction, FL
behavior is observed at low temperatures $T$
with a specific heat $c \propto T$, a magnetic
susceptibility $\chi \approx const.$ and a resistivity $\rho \approx
\rho_0+A T^2$.
However, in several heavy-fermion systems
\cite{seaman,andraka1,maple,bernal,loehneysen1,tuning,loehneysenUeb} 
pronounced
deviations from FL behavior have been found in a number of
physical properties.

 Three
main theoretical scenarios have been proposed to explain the
occurrence of the non-Fermi-liquid (NFL) behavior:
 In the first one \cite{milanovic,langenfeld,bernal} it is assumed
that disorder introduces
a distribution of (one-channel) Kondo temperatures $T_K$ in the system; this
distribution can be directly related to an anomalous low temperature
behavior.  The second model proposes a single-impurity origin of the
NFL, e.g. associated with the
 quadrupolar (two-channel) Kondo
effect \cite{seaman,maple}.

\begin{figure}
 \epsfig{width=0.9 \linewidth,file=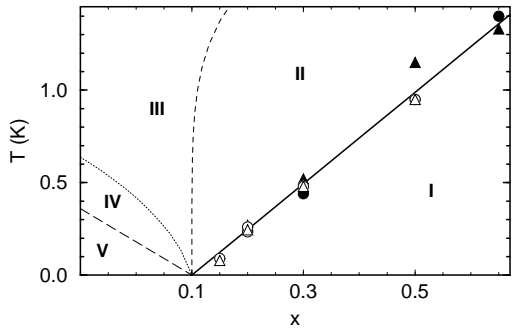}
 \caption{Phase diagram of CeCu$_{6-x}$Au$_x$ as a function 
   of doping and temperature.  The points are 
   N\'eel temperatures determined by different methods \citeL\ 
(open and closed symbols for single and polycrystals respectively), 
the solid line denotes the phase transition,
the dashed lines are theoretical crossover lines. The regions
 are described in the text.}
 \label{phasediagram.fig}
\end{figure}

In CeCu$_{6-x}$Au$_x$, there is clear experimental evidence 
\cite{loehneysen1,tuning,loehneysenUeb} for a
third mechanism based on the proximity to a
 quantum phase transition (QPT) \cite{hertz,millis,sachdev,continentino} 
near $x_c=0.1$. 
For $x > x_c$ the system
orders magnetically with a N\'eel temperature 
$T_N \propto (x-x_c)^\mu$ with $\mu=1$ as shown 
in Fig. \ref{phasediagram.fig}.
The QPT can be interpreted as the result of the competition between
the Kondo effect, which tends to screen the magnetic moments, and
the RKKY interaction, which favors a magnetically ordered state.
The Kondo effect is weakened by increasing doping which, as
suggested experimentally, only leads to a volume increase with no apparent
change in the number of carriers.

\begin{figure}
 \epsfig{width=0.95 \linewidth,file=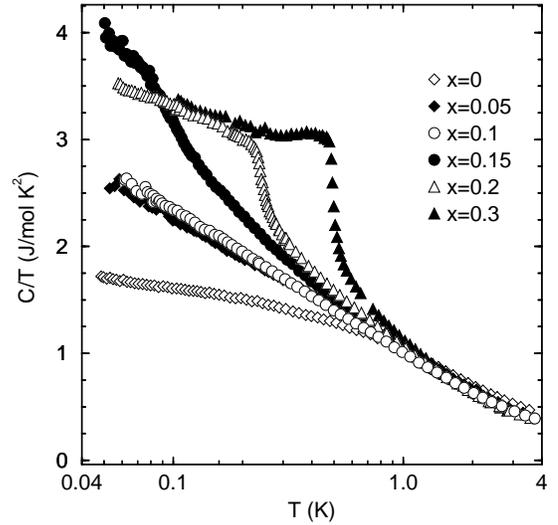}
 \caption{The specific heat $c/T$ of  CeCu$_{6-x}$Au$_x$ versus 
   $\log T$  (from \citeL).}
 \label{specificHeat.fig}
\end{figure}

CeCu$_{6-x}$Au$_x$ remains metallic for all $x$ with typical
FL properties at low temperatures 
both at $x=0$ and for $x\gg x_c$. However, at
$x=x_c$, $T_N$ vanishes and NFL behavior is
observed in all accessible quantities down to the lowest 
temperatures, $T \ll T_K\approx 6 K$.
 As an example we show in Fig.
\ref{specificHeat.fig} the specific heat with $c/T \propto \ln
(T_0/T)$ over nearly two decades. At the same doping, the
static susceptibility $\chi \propto 1-\alpha \sqrt{T}$ and the
resistivity $\rho\approx \rho_0+A' T$ show a remarkable NFL 
behavior over a substantial $T$ range \cite{loehneysen1}.

Up to now \cite{loehneysen1,loehneysenUeb,moriya} it was assumed that
the critical fluctuations of the QPT are
dominated by incommensurate three-dimensional correlations. This
would strongly suggest a description by a quantum-critical theory with
$d=3, z=2$ as investigated in \cite{hertz,millis,moriya}.  However,
this well-established 
theory is in contradiction to the
thermodynamic properties which are observed in the experiment, as it
would suggest $c/T \propto 1-B \sqrt{T}$ and $\rho \approx \rho_0+
A'' T^{3/2}$ at low temperatures \cite{millis,moriya}.  One could,
however, argue that in the experiments only a crossover region is
accessible \cite{moriya}. In our opinion this is not fully
convincing and it appears not to be possible to fit both resistivity and
specific heat over the observed $T$ 
range by such a theory \cite{loehneysenUeb,resistivity}.

\begin{figure}
 \epsfig{width=0.8 \linewidth,file=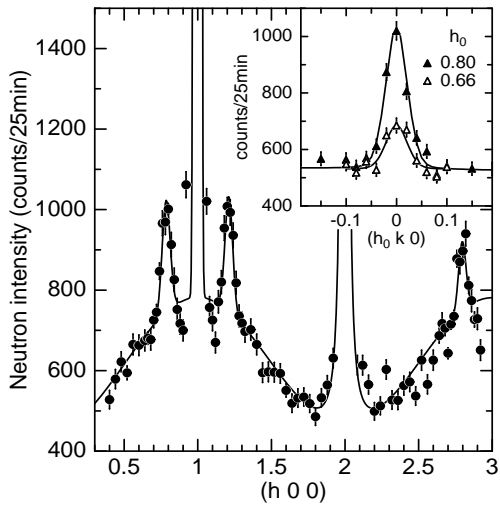}
 \caption{Elastic neutron scattering of CeCu$_{5.8}$Au$_{0.2}$ 
 along the $a$ axis 
 at $70$mK. The inset shows that all features are sharp 
 in the $b$ direction.}
 \label{elastic.fig}
\end{figure}

Below we will focus on a novel feature of the magnetic fluctuations which
we uncovered by carrying out elastic neutron scattering experiments near
the quantum critical point (QCP). 
The data taken in the ordered phase ($x=0.2$) is displayed in 
Fig. \ref{elastic.fig}. A scan
along $(h,0,0)$ 
reveals besides the nuclear reflections two magnetic features which
vanish for higher temperatures. 
The satellite peaks at $(\pm 0.79,0,0)$ describe
a three-dimensional incommensurate magnetic order. The magnitude of
the ordered magnetic moment is extremely small and is estimated  
to be of the order of $0.02 \mu_B$. 
The correlation length determined from  the peak width
is approximately $20$ 
lattice constants $a$ (orthorhombic notation), with $4$ Ce atoms per unit cell.

The second feature, seen as a
 broad  sine-modulated background with maxima at 
(1, 0, 0) and (3, 0, 0) in Fig. \ref{elastic.fig},
has a factor of $3-4$ higher integrated weight. It is important
to stress 
that both features are sharp in the
$b$ direction with a width of order $20$ unit cells as can be seen
from scans along  $(h_0, k, 0)$ for different $h_0$ (Fig. \ref{elastic.fig}).

In this letter we propose that the magnetic fluctuations associated with
the second broad structure in the neutron scattering 
dominate the critical
fluctuations at the QPT in the observed temperature range.
 Starting from this
assumption we will show that the logarithmic increase of $c/T$ for decreasing
temperature,
 the linear resistivity and the phase diagram can easily be
derived.  We interpret this structure as arising from
ferromagnetically ordered
planes perpendicular to the $a$ direction. The sine modulation
would suggest that effectively two of these ferromagnetic planes with
distance $a/2$ couple
antiferromagnetically in the $a$ direction, but different pairs of planes
are incoherent. While it is possible to identify slightly corrugated
planes of Ce atoms in the crystal structure, we do not see an obvious
reason for a strongly asymmetric coupling, which could directly
explain the observed two-dimensional structure.
 For the following, the details of the magnetic order
are irrelevant and we will  only assume that some critical two-dimensional 
fluctuations exist.

To derive an effective action  near the phase transition 
in a Ginzburg-Landau-Wilson  approach
\cite{hertz} we have to account for
the damping of the critical fluctuations. To describe the critical
modes, we introduce a scalar field $\Phi_{{\bf q}_{||}}$; the
experiments suggest a preferred direction of the magnetic moments along
the $c$ axis \cite{loehneysenUeb,almut}.  ${\bf q}_{||}$ is a
two-dimensional vector in the $bc$ plane. Ordering will occur at ${\bf
  q}_{||}=0$. The primary damping mechanism is the coupling to
particle-hole pairs. The dynamics of the quasi-particles is
 three-dimensional, this can be
  inferred from the transport properties which  vary at most  by a
factor of two in different directions \cite{resistivity}. 
We assume a coupling of the
critical fluctuations to the heavy quasi-particles (with creation operators
$c^\dagger_{\bf k}$) by the following Hamiltonian:
\begin{equation} \label{H}
H_c= g \sum_{{\bf k},{\bf q}, \alpha, \beta}
\left( c^\dagger_{\alpha,{\bf k}+{\bf q}} \sigma^z_{\alpha \beta} 
c_{\beta,{\bf k}}  \right)
\Phi_{{\bf q}_{||}} h(q_{\perp}).
\end{equation}
${\bf q}={\bf q}_{||}+ {\bf q}_{\perp}$ is the transferred momentum
split up in two components parallel and perpendicular to the
planes. $\sigma^i$ are the Pauli matrices and $g$ is the
coupling constant.  $h(q_{\perp})$ is some smooth function describing the
magnetic structure perpendicular to the planes, e.g. $h(q_{\perp})=i
\sin q_{\perp} a/4$ for two planes with distance $a/2$ which are
 coupled antiferromagnetically. The details of $H_c$
are, however, not important for our discussion.

Integrating out the fermions (cf. \cite{hertz}
) induces damping of and interaction between the critical modes
$\Phi_{{\bf q}_{||},\omega}$. To quadratic order,
 we obtain the following contribution to the 
effective action in imaginary time ($\beta=1/(k_B T)$):
\begin{displaymath}
S_d=\frac{g^2}{\beta} \sum_{\omega_n, {\bf q}_{||}}  
\left(\int   |h(q_\perp)|^2 \chi^0({\bf q},i \omega_n) d{q_{\perp}} \right)
 \left|\Phi_{{\bf q}_{||},i \omega_n}\right|^2.
\end{displaymath}
$\omega_n=2 \pi n/\beta$ are bosonic Matsubara frequencies.  The
damping is described by the imaginary part of the particle-hole bubble
$\chi^0$, $\text{Im} \int d q_{\perp} |h(q_\perp)|^2 \chi^0({\bf
  q},\omega+i 0)\approx \gamma \omega
$. 
Near the QCP, after a proper rescaling,
 the effective action takes
 the following form:
\begin{eqnarray} \label{S}
S&=&S_2+S_{\text{int}} \\
S_2&=&\frac{1}{\beta}  \sum_{\omega_n}\!  \int 
\Phi^{*}_{{\bf q}_{||},i \omega_n} \left(\delta+{\bf q}_{||}^2+
|\omega_n| \right) \Phi_{{\bf q}_{||},i \omega_n} 
 d^2 {\bf q}_{||}. \nonumber
\end{eqnarray}
$S_{\text{int}}$ describes the interaction between the critical
modes, the leading term is given by $S_{\text{int}}
\approx U \int_0^\beta \! d \tau \int\! d^2{\bf r} |\Phi({\bf r},
\tau)|^4$ \cite{hertz}. The distance from the QCP is
measured by $\delta$. At a critical value $\delta_c$, determined by the
strength of interaction, the system is at the QCP,
i.e.  $\delta-\delta_c \propto x_c-x$.
The effective action corresponds to a quantum-critical theory in $d=2$
with a dynamic exponent $z=2$. This dynamic exponent describes the
fact that if one scales momenta by ${\bf k} \to \lambda {\bf k}$ one
has to scale frequencies 
by $\omega \to
\lambda^z \omega$. 

This theory has been widely investigated by many
authors (e.g. \cite{hertz,millis,sachdev}), mainly in the context
of the antiferromagnetic spin -fluctuation picture of 
high-$T_c$ compounds. We will primarily
employ the results of Millis \cite{millis} who did a careful
renormalization-group study of Eq. (\ref{S}). The effective dimension of
this theory at $T=0$ is $d+z=4$; the scaling dimension of the
interaction $S_{\text{int}}$ is $4-(d+z)=0$, therefore it is marginal.
Nevertheless, the leading behavior of the specific heat in the
disordered phase can be directly calculated from the Gaussian part,
$S_2$, of the action of Eq. (\ref{S}). The free energy per volume
corresponding to $S_2$ is
\begin{eqnarray}
F&=&\int_0^{\Lambda_k} \frac{d^2 {\bf q}_{||}}{(2 \pi)^2} 
\int_0^{\Lambda_\omega}
\frac{d \epsilon}{\pi} \coth \frac{\epsilon}{2 T} 
\arctan \frac{\epsilon}{\delta+{\bf q}_{||}^2} \nonumber \\
&\propto&  \frac{T^2}{\omega_c} \ln \frac{\omega_c^2}{T^2+
(\delta-\delta_c)^2}.
\end{eqnarray}
Note that $\delta_c=0$ in this approximation; 
$\omega_c$ is a typical cutoff energy of the order of the
Kondo temperature in the system. Consequently, the coefficient of
the specific heat $\gamma=c/T=-d^2 F/d T^2$ diverges logarithmically
with decreasing temperature at the QCP, as observed in the
experiment. For $\delta>\delta_c$, i.e. $x<x_c$, $\gamma$ stays
finite. Note that the logarithmic terms do not arise from some
marginal operators, they are due to a pure phase space effect, this typically
happens at $d+z=2 z$ \cite{millis}.

By a solution \cite{millis} of the scaling equations for the model of
Eq. (\ref{S}) a phase diagram emerges as displayed 
in Fig. \ref{phasediagram.fig}.  
Region I, the low temperature phase for
$x>x_c$, is the ordered phase.  The behavior near $T_N$
 will depend on the structure of the order
parameter, e.g., whether the ordering is of Ising type or how it is
stabilized by three-dimensional coherence. The Ginzburg criterion
for our model in Eq. (\ref{S}) predicts \cite{millis} that $T_N$
is proportional to the distance from the critical
point, $T_N \propto x-x_c$. This relation is fulfilled by the
experiment over an astonishingly large range. The other
crossover lines are hard to analyze quantitatively with the existing
experimental
data, therefore we do not attempt to fit them. The qualitative trend
is, however, consistent with the phase diagram. In Fig.
\ref{phasediagram.fig} we include the theoretical curves of
\cite{millis}.  Region II is dominated by the fluctuations of the
finite-temperature phase transition; in region III true
quantum-critical behavior with $\gamma \propto \ln (T_0/T)$ and, as we
shall see, $\rho \approx \rho_0+A' T$ can be observed.  In this 
quantum-critical regime -- in \cite{millis} it is called ``classical Gaussian
regime'' -- the temperature is the most important energy scale. The
correlation length $\xi$ depends on temperature as $\xi^{-2}
\propto T$ with logarithmic corrections.  In region IV, a pure
crossover regime which is probably hard to observe, $\xi$ is
determined by the distance from the critical point, $\xi^{-2} \propto
x_c-x$, while energy fluctuations and therefore the specific heat are
governed by the temperature with $\gamma \propto \ln (T_0/T)$. For region
V, FL behavior is expected with a linear specific heat
$c/T \propto -\ln (x_c-x)$ and a finite correlation length $\xi^{-2}
\propto x_c-x$ as in region IV.

The whole scenario is in qualitative agreement with the experiments
\cite{loehneysen1,tuning,loehneysenUeb}. In particular,
the logarithmic increase of the specific-heat coefficient
in region III, the $T$-linear
resistivity which we will calculate in the following, and the
linear increase of $T_N$ with $x$ are confirmed by the experiments.

For the calculation of the resistivity we closely follow Hlubina and
Rice \cite{hlubina} who have calculated the resistivity of electrons
in two dimensions coupled to quantum-critical antiferromagnetic
spin fluctuations. In our case the dynamics of the quasi-particles
is  three-dimensional 
as mentioned above.  Our starting point
is Eq. (\ref{H}) and the corresponding collision term in a Boltzmann
equation:
\begin{eqnarray}\label{collision}
\left. \frac{\partial f_{{\bf k}}}{\partial t}\right|_{\text{coll}}
&=&\frac{2 g^2}{T} 
\sum_{{\bf k}'}
\int_{-\infty}^\infty d \omega n(\omega) f^0_{{\bf k}'} (1-f^0_{{\bf k}}) \\
&& \times (\varphi_{{\bf k}'}-\varphi_{{\bf k}}) 
\delta(\epsilon_{{\bf k}}-\epsilon_{{\bf k}'}-\omega) 
\text{Im} \chi_{{\bf k}'_{||}-{\bf k}_{||}}(\omega).\nonumber
\end{eqnarray}
Here we have already linearized the collision term; the occupation of
a state with momentum ${\bf k}$ is given by $f_{{\bf k}}=f^0_{{\bf
    k}}+\varphi_{{\bf k}} (\partial f^0_{{\bf k}}/\partial \epsilon)$
where $f^0_{\bf k}=f^0(\epsilon_{\bf k})$ is the usual Fermi function;
$n(\epsilon)$ is the Bose function. We have omitted the factor
$|h(q_\perp)|^2$, the qualitative behavior of the resistivity is not
influenced by this smooth function.
$\chi_{\bf q}(\omega)=
\left\langle \Phi^*_{{\bf q},\omega} \Phi_{{\bf q},\omega}
\right\rangle$ is the order-parameter susceptibility. In the
following we use for the disordered phase
\begin{eqnarray}\label{chi}
\chi_{\bf{q}_{||}}(\omega)\approx\frac{A}{T^*+c T+{\bf q}_{||}^2-i \omega}
\end{eqnarray}
with temperature-independent constants $A,c$ and $T^*$. This 
phenomenological form corresponds to the behavior of the correlation 
length described above,
 $\xi^{-2}\propto T$ in the quantum-critical region up to
logarithmic corrections.  $T^* \propto \delta-\delta_c$ measures the
distance from the QCP. Eq. (\ref{chi}) is valid in
regions III-V, we are primarily interested in the quantum-critical regime
III with $T^* \ll T$. Following \cite{hlubina},
 the resistivity
can be determined within Boltzmann theory from the minimization of a 
functional of $\varphi_{{\bf k}}$
\begin{eqnarray}\label{rho}
\frac{\rho}{\rho_0}=\min_{\varphi_{\bf k}} \left[ 
\frac{\sum_{{\bf k} {\bf k}'}
 W_{{\bf k} {\bf k}'} (\varphi_{{\bf k}}-\varphi_{{\bf k}'})^2}
{\left\{\sum_{{\bf k}} \varphi_{{\bf k}} {\bf v}_{{\bf k}} \cdot \hat{n} 
(-\partial f^0_{{\bf k}}/\partial \epsilon) \right\}^2    }  \right],
\end{eqnarray}
where $\hat{n}$ is the direction of the applied electric field,
${\bf v}_{{\bf k}}$ the velocity of the electrons, $\rho_0=\hbar/e^2$ and
$$
W_{{\bf k} {\bf k}'}=\frac{2 g^2}{T} f^0_{{\bf k}} (1-f^0_{{\bf
    k}'}) n(\epsilon_{{\bf k}'}-\epsilon_{{\bf k}}) \text{Im}
\chi_{{\bf k}'_{||}-{\bf k}_{||}} (\epsilon_{{\bf k}'}-\epsilon_{{\bf
    k}}).
$$
Note that a second contribution due to impurity scattering has to
be added, which is not given here.  For simplicity we assume 
a spherical Fermi surface. The
radial part of the momentum integration can be rewritten as an energy
integration. Using $\int f^0(\epsilon)(1-f^0(\epsilon+\omega))
d\epsilon= \omega (1+n(\omega))$ and $\int_0^{\infty} \omega n(\omega)
[1+n(\omega)] \text{Im} \chi_{{\bf k}'_{||}-{\bf k}_{||}}(\omega) d
\omega = I[\beta (T^*+c T+({\bf k}'_{||}-{\bf k}_{||})^2)] T A$ with
$I[x]\approx \pi^2/(3 x (x+2 \pi/3))$ \cite{hlubina} we can perform the
energy integration.
As long as the resistivity is dominated by impurity scattering -- this
is true in the whole range where a linear temperature dependence has
been observed -- we can assume \cite{hlubina} $\varphi_{\bf k}\approx
{\bf v}_{\bf k} \cdot \hat{n}$ and arrive at
\begin{eqnarray}
\Delta \rho &\propto& T^2 \int \frac{g(\alpha)}{(T^*+c T+k_F^2 \alpha) 
(T^*+c' T+k_F^2 \alpha)} d \alpha \\
g(\alpha)&=& \int \!\!\! \int  d \Omega_{\bf k}  d \Omega_{{\bf k}'} 
(\hat{n} \cdot \hat{\bf k}-\hat{n} \cdot \hat{{ \bf k}}')^2 
\delta(\alpha-(\hat{{\bf k}}'_{||}-\hat{\bf k}_{||})^2) \nonumber
\end{eqnarray}
with $c'=c+2 \pi/3$. $\hat{\bf k}$ denotes a unit vector in the
 direction of ${\bf k}$ and $\hat{\bf k}_{||}$ its
projection to the $bc$ plane.
For $\hat{n}$ not parallel to the planes, we find $g(\alpha)\approx
const$ for small $\alpha$ and therefore we get approximately
\begin{eqnarray}
\Delta \rho \propto \frac{T}{c-c'} \ln\frac{T^*+c T}{T^*+c' T} 
\approx \left\{ \begin{array}{ll} T^2 &   , T<T^*/c' \\
 T {\displaystyle \frac{\ln c/c'}{c-c'}} &  , T>T^*/c
\end{array} \right. \!\!.
\end{eqnarray}
For a finite $T^*$ and small temperatures we recover the usual $\Delta
\rho \propto T^2$ of a FL. However, in the quantum critical regime, 
i.e. for $T^* \ll c T$, the resistivity is linear in temperature as
observed in the experiment \cite{loehneysen1,loehneysenUeb,resistivity}.

For an electric field parallel to the planes, we obtain
$g(\alpha)\propto -\alpha \ln \alpha$ and accordingly $\Delta \rho
\propto T^2 \ln T$ for $T^*=0$. This is however an artifact of our
approximation of using a spherical Fermi surface and an isotropic $s$-wave
scattering amplitude.
  In a more realistic approach, different directions would
mix and give a linear increase of resistivity with temperature in all
directions, as observed \cite{resistivity}.  Nevertheless, one would
still expect that the linear increase of the resistivity at the
QCP is largest in the direction perpendicular to
the plane which is indeed observed \cite{resistivity}.

The NFL character is also manifest in an anomalous self-energy of
the electrons. Calculation of the lifetime in Born approximation at $T=0$
using Eqs. (\ref{H}) and (\ref{chi}) gives  
$1/\tau_{\bf k} \propto \epsilon_{\bf k} \ln 1/\epsilon_{\bf k}$ 
for directions parallel to the planes and 
$1/\tau_{\bf k} \propto \epsilon_{\bf k}$ in the $a$ direction. 
Averaging over the Fermi surface results in 
$\left\langle 1/\tau_{\bf k} \right\rangle \propto \epsilon_{\bf k} 
\ln 1/\epsilon_{\bf k}$, in sharp contrast to the usual $1/\tau_{\bf k} 
\propto \epsilon_{\bf k}^2$ in a Fermi liquid.

In \cite{loehneysen1,loehneysenUeb} the static susceptibility for $x=x_c$
was fitted by $\chi \propto 1-\alpha \sqrt{T}$ from the lowest measuring
temperature of $80$ mK up to $3$ K. However, the data can also 
be well described by 
$\chi \approx a_0+1/(a_1+a_2 T)$ for temperatures up to $1.4$ K.
One would expect a susceptibility of the last form for 
two antiferromagnetically coupled planes with an 
order-parameter susceptibility as given by Eq. (\ref{chi}). Further
theoretical and experimental studies are needed to clarify
this point. It will also be important to investigate further the
interplay of the two- and three-dimensional order
which could finally lead to a change in the observed properties
at some lower temperature. We think that the explanation
of the phase diagram, of the linear increase of the resistivity with
temperature and of the anomalous specific heat in CeCu$_{6-x}$Au$_x$
by a quantum-phase transition with $d=2, z=2$ is already promising.
To our knowledge, this would establish the first clear experimental 
realization of such a theoretical scenario.

The authors wish to thank A. Ruckenstein, P. W\"{o}lfle and T. Pietrus
for 
many stimulating discussions. 
The neutron scattering experiments were performed at 
the Ris\o{} National Laboratory,
Denmark. This work was supported by the DFG and LIP.


\begin{thebibliography}{99}
\vspace{-1.2cm}
\bibitem{seaman} C. L.
  Seaman {\it et al.}, Phys. Rev.  Lett.  {\bf 67}, 2882 (1991).
\bibitem{andraka1} B. Andraka and A.  M. Tsvelik, Phys. Rev. Lett.
  {\bf 67}, 2886 (1991); B. Andraka and G. R.
  Stewart, Phys. Rev. B {\bf 47}, 3208 (1993).
\bibitem{maple} M. B. Maple {\it et al.}, J.
  Low. Temp. Phys. {\bf 95}, 225 (1994).  
\bibitem{bernal} O. O. Bernal {\it et al.},  Phys. Rev. Lett.
  {\bf 75}, 2023 (1995).
\bibitem{loehneysen1} H. v.  L\"{o}hneysen {\it et al.},
 Phys. Rev.  Lett.  {\bf 72}, 3262 (1994).  
\bibitem{tuning} B.
  Bogenberger and H. v. L\"{o}hneysen, Phys. Rev. Lett.  {\bf 74},
  1016 (1995).  
\bibitem{loehneysenUeb} H. v.  L\"{o}hneysen, Physica
  B {\bf 206} \& {\bf 207}, 101 (1995); H. v.  L\"{o}hneysen, 
J. Phys. Cond. Matt. {\bf
    8}, 9689 (1996).  
\bibitem{milanovic} M. Milanovi\'c, S. Sachdev, and R. N. Bhatt, 
Phys. Rev. Lett.
  {\bf 63}, 82 (1989); V. Dobrosavlejevi\'c, T. R. Kirkpatrick, 
and G. Kotliar,
   Phys. Rev. Lett.
  {\bf 69}, 1113 (1992).
\bibitem{langenfeld} A. Langenfeld and P. W\"{o}lfle, 
  Ann. Phys. {\bf 4}, 43 (1995).
\bibitem{hertz} J. A. Hertz, Phys. Rev. B {\bf 14}, 1165
  (1976).  
\bibitem{millis} A. J. Millis, Phys. Rev. B {\bf 48}, 7183
  (1993).  
\bibitem{sachdev} S. Sachdev, A. V. Chubukov, and A. Sokol, Phys. Rev. B
  {\bf 51}, 14874, (1995).  
\bibitem{continentino} M. A. Continentino, Z. Phys. B {\bf 101}, 197 (1996). 
\bibitem{moriya} T. Moriya and
  T. Takimoto, J. Phys.  Soc. Japan {\bf 64}, 960 (1995); 
S. Kambe {\it et al.},
{\it ibid.} {\bf 65}, 3294 (1996).  
\bibitem{resistivity} A. Neubert  {\it et al.}, Physica B (in print).  
\bibitem{almut} A. Schr\"{o}der {\it et al.}, 
Physica B {\bf 199}  \& {\bf 200}, 47 (1994). 
\bibitem{hlubina} R. Hlubina and T. M. Rice, Phys. Rev. B
  {\bf 51}, 9253 (1995).  
\end{thebibliography}
\end{document}